\documentstyle[prb,aps,multicol,epsfig]{revtex}

\begin{document}

\title{Observation of local chiral-symmetry breaking in globally centrosymmetric crystals.}

\author{S. Di Matteo$^{a,b}$, Y. Joly$^{c}$, A. Bombardi$^{d}$, L. Paolasini$^{d}$, F. de Bergevin$^{d}$, C.R. Natoli$^{a}$}

\address{$^a$INFN Laboratori Nazionali di Frascati, c.p. 13, I-00044 Frascati, Italy \\
$^b$INFM Udr Roma III, via della Vasca Navale 142, 00100 Roma, Italy \\
$^c$Laboratoire de Cristallographie, CNRS, B.P. 166, F-38042, Grenoble
Cedex 9, France \\
$^d$E.S.R.F., B. P. 220, F-38043, Grenoble Cedex 9, France}

\date{\today}

\maketitle

\begin{abstract}
A thorough tensor analysis of the Bragg-forbidden reflection (00.3)$_h$ in corundum systems having a global center of inversion, like V$_2$O$_3$ and  $\alpha$-Fe$_2$O$_3$, shows that anomalous x-ray resonant diffraction can access chiral properties related to the dipole-quadrupole (E1-E2) channel via an interference with the pure quadrupole-quadrupole (E2-E2) process. This is also confirmed by independent {\it ab initio} numerical simulations. In such a way it becomes possible, in this particular case, to estimate the intensity of the ``twisted'' trigonal crystal field ($C_3$ symmetry) and, in general, to detect chiral quantities in systems where dichroic absorption techniques are ineffective.
\end{abstract}

\begin{multicols}{2}


In recent years third generation radiation sources have made possible the detection of relatively small effects in crystal electronic structure due either to magnetic anisotropy\cite{paolasini} or to interference between dipole and quadrupole (E1-E2) transition matrix elements. \cite{goulon,sawatzky} For example, it is now well established that $K$-edge circular dichroism in absorption is sensitive to the angular orbital moment $L_z $ in magnetic spectroscopies like XMCD (x-ray magnetic circular dichroism), or to the peculiar physical quantity $L_z \Omega_z$ in non-magnetic spectroscopies like  XNCD \cite{goulon,benoist} (x-ray natural circular dichroism). Here $\vec{\Omega} = \hat{r}\wedge\vec{L} - \vec{L}\wedge\hat{r}$ is the toroidal (anapole\cite{benoist}) orbital moment, and $z$ is the direction of the incoming photon. 
As apparent from the definition of  $\vec{\Omega}$, the product $L_z\Omega_z$ is time-reversal even and inversion odd. Therefore if a paramagnetic system has a global inversion symmetry, even though at a local level this symmetry is broken, the resulting XNCD signal vanishes. A similar situation arises, {\it mutatis mutandis}, for XMCD in many antiferromagnets, where the locally broken time-reversal symmetry can be globally restored, making the total dichroic signal zero.

A common way to circumvent this limitation is to use anomalous x-ray diffraction, where the local transition amplitudes are added with a phase factor that can compensate the vanishing effect due to the global symmetry. This technique has been widely used for antiferromagnets to study local magnetic effects.\cite{paolasini} 
However to our knowledge no such  technique has been reported to measure the quantity $L_z\Omega_z$ for systems with a global inversion symmetry. 

It is the purpose of this paper to show that this is possible, by presenting a comprehensive theoretical analysis of the measured Bragg forbidden (00.3)$_h$ ``Finkelstein''  reflection,\cite{finkel} in systems belonging to the corundum crystal class having a global inversion symmetry, namely V$_2$O$_3$ \cite{paolasini2} and $\alpha$-Fe$_2$O$_3$.\cite{finkel} In this latter compound, for example, this reflection has been interpreted\cite{finkel,carra} as a pure E2-E2 process, giving rise to a six-fold periodicity in the azimuthal scan around the trigonal axis. We show here that such a description is not complete, as a non negligible E1-E2 term is present, in V$_2$O$_3$ as well as in  $\alpha$-Fe$_2$O$_3$ and actually in all the corundum systems. Its presence gives rise to a three-fold modulation of the azimuthal scan that is recognizable in the spectra of Paolasini {\it et al.}\cite{paolasini2}, appears much more pronounced in Finkelstein {\it et al.}\cite{finkel} (see Figs. 2 of both papers) and is related to the operator $L_z\Omega_z$, as shown below. This modulation is also present in a more recent experiment by K. Ishida {\it et al.}\cite{ishida} on  $\alpha$-Fe$_2$O$_3$.

\begin{figure}
\centerline{\epsfig{file=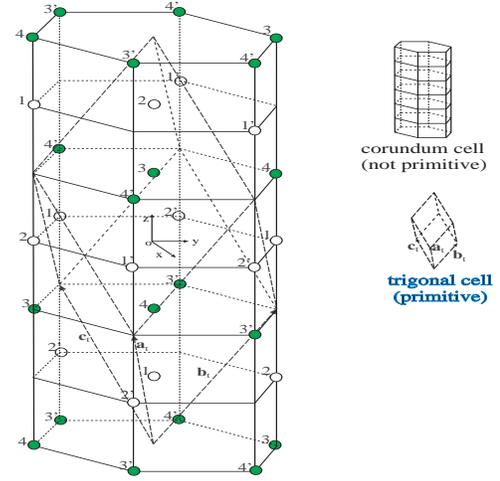,height=2.5in,width=2.5in}}
\vspace{5pt}
\caption
{Corundum and trigonal (rhombohedral) cells.}
\label{cell}
\end{figure}

\noindent 
We first focus our theoretical analysis on the paramagnetic corundum phase of V$_2$O$_3$ (space group R${\overline {3}}$c), so that we can neglect magnetic contributions. We shall comment on the general magnetic case later on. The four vanadium ions of the unit cell lie along the trigonal axis (see Fig. \ref{cell}),  slightly offset from the centre of the distorted oxygen octahedra.  Their positions in rhombohedral coordinates are\cite{dernier}:  $V_1 = (t, t, t), \; V_2 = (1-t, 1-t, 1-t), \; V_3 = (1/2+t, 1/2+t, 1/2 +t), \; V_4 = (1/2-t, 1/2 -t, 1/2-t)$, with t=0.1537. The four metallic sites are related by the following symmetry operations:  V$_2$=${\hat I}$V$_1$, V$_3$=${\hat m}_x$V$_1$, V$_4$=${\hat C}_{2x}$V$_1$, where ${\hat I}$  is the inversion in the reference frame of Fig. \ref{cell} and ${\hat C}_{2x}$  and  ${\hat m}_x$ are, respectively, the two-fold rotation around $x$-axis and the mirror plane orthogonal to it.


In the resonant regime the $K$-edge elastic scattering amplitude, $A({\vec q},\omega)$, can be  written in terms of the atomic scattering factors (ASF), $f_j(\omega)$, of the atoms at positions $\rho_j$ in the unit cell as:

\begin{equation}
A(\vec{q},\omega) = \Sigma _j e^{i\vec{q} \cdot \vec{\rho_j}} f_j(\omega)
\label{bragg}
\end{equation}

\noindent where $\hbar \vec{q}$ is the momentum transfer in the scattering process, $\hbar \omega$ the incoming and outgoing photon energy and the sum is over the four atoms in the rhombohedral primitive cell. The ASF is a second-order process in the electron-radiation interaction, whose explicit expression for core resonances reads, in atomic units:

\begin{equation}
f_j(\omega) = (\hbar \omega)^2 \sum _n \frac{\langle \Psi_0^{(j)}|{\hat{O}}^{+}_s|\Psi_n \rangle \langle \Psi_n|{\hat {O}_i}|\Psi_0^{(j)}\rangle}
{\hbar \omega - (E_n-E_0) -i\Gamma_n}
\label{asf}
\end{equation}

\noindent where $|\Psi_0^{(j)}\rangle$ is the ground state, with the origin taken on the $j$-th scattering atom, and  $E_0$ its energy; the sum is over all the excited states $|\Psi_n \rangle$, with corresponding energies $E_n$. Finally $\Gamma_n$ is a damping term that takes into account the core-hole and the finite life-time of the excited states $|\Psi_n \rangle$ and the indices $i, (s)$ refer to the incident (scattered) properties of the photon field.
Neglecting the electric dipole-magnetic dipole terms, which are usually small in the x-ray range, we can write the operator ${\hat {O}}$ up to the electric quadrupole contribution as:

\begin{equation}
{\hat {O}}= \vec{\epsilon}\cdot\vec{r} (1-\frac{i}{2}\vec{k}\cdot\vec{r})
\label{transop}
\end{equation}

\noindent where $\vec{\epsilon}$ and $\vec{k}$ are the polarization and wave vector of the x-ray photon and $r$ is the coordinate of the electron in the reference frame of the scattering atom. The matrix element in Eq. (\ref{asf}) depends only on the electronic part of the operator ${\hat {O}}$, in such a way that the radiation parameters, $\vec{\epsilon}$ and $\vec{k}$, can be factorised. After some algebra Eq. (\ref{asf}) can be written as a scalar product of two irreducible tensors\cite{carra,natoli}:

\begin{equation}
f_j(\omega)= \Sigma _{p,q} (-)^q T^{(p)}_q F^{(p)}_{-q}(j;\omega)
\label{asffinal}
\end{equation}

\noindent where $T^{(p)}_q$ depends on the incident and scattered polarizations and wave vectors and $F^{(p)}_q$ is the tensor representing the properties of the system. It is important to note that $F^{(p)}_q (j;\omega) \equiv \langle \Psi_0^{(j)}| {\hat {F}}^{(p)}_q(\omega) |\Psi_0^{(j)}\rangle$ is such that it must belong to the totally symmetric representation of the local point group of the scattering atom (the $A_{1g}$ representation, in Bethe's notation). In fact it is easy to check that if  ${\hat {R}}$ is a symmetry operation for $|\Psi_0^{(j)}\rangle$ (ie, ${\hat {R}}|\Psi_0^{(j)}\rangle = |\Psi_0^{(j)}\rangle$), then

\begin{equation}
\langle \Psi_0^{(j)}| {\hat {F}}^{(p)}(\omega) |\Psi_0^{(j)}\rangle = \langle \Psi_0^{(j)}| {\hat {R}}^{-1}{\hat {F}}^{(p)}(\omega){\hat {R}} |\Psi_0^{(j)}\rangle
\label{totsym}
\end{equation}

\noindent  Thus, the only allowed  matrix elements are those of the components of ${\hat {F}}^{(p)}(\omega)$ that are invariant for any symmetry elements of the point group.

In the corundum systems the local symmetry on each scattering atom is ${\hat {C}}_3$. Thus Eq. (\ref{totsym}) imposes the following restriction: only the irreducible tensors whose azimuthal numbers with respect to the three-fold axis are 0 or $\pm$3 are allowed. A similar constraint was recognized for the case of haematite,\cite{carra} although the analysis was there limited to an octahedral symmetry, loosing therefore the E1-E2 contribution. If we choose the quantization axis coincident with the three-fold axis, the tensor components allowed by the ${\hat {C}}_3$ symmetry are:

a) in the E1-E1 channel: $F^{(0)}_0$, ${\tilde {F}}^{(1)}_0$ and $F^{(2)}_0$

b) in the E1-E2 channel: $F^{(1)}_0$, $F^{(2)}_0$, $F^{(3)}_0$, $F^{(3)}_3 \pm F^{(3)}_{-3}$,
${\tilde {F}}^{(1)}_0$, ${\tilde {F}}^{(2)}_0$, ${\tilde {F}}^{(3)}_0$, ${\tilde {F}}^{(3)}_3 \pm {\tilde {F}}^{(3)}_{-3}$

c) in the E2-E2 channel: $F^{(0)}_0$, ${\tilde {F}}^{(1)}_0$, $F^{(2)}_0$, ${\tilde {F}}^{(3)}_0$, $F^{(4)}_0$, ${\tilde {F}}^{(3)}_3 \pm {\tilde {F}}^{(3)}_{-3}$ and $F^{(4)}_3 \pm F^{(4)}_{-3}$

The tilded tensors refer to time-reversal odd (ie, magnetic) quantities. As V$_2$O$_3$ in the corundum phase is paramagnetic, tilded tensors are all zero. Another extinction rule comes from the structure factor. We are interested in the (00.3)$_h$ reflection, (111)$_r$ in the rhombohedral system. For $\vec{q}$=(111)$_r$,  Eq. (\ref{bragg}) becomes:

\begin{eqnarray}
A(\vec{q},\omega) = e^{2\pi i t} f_1+e^{-2\pi i t} f_2+e^{i \pi}e^{2\pi i t} f_3 \nonumber \\
+e^{i \pi}e^{-2\pi i t} f_4 = (e^{2\pi i t}+e^{-2\pi i t}{\hat {I}})(1-{\hat {m}}_{x}) f_1  \nonumber \\
= \left\{ \begin{array}{ll}
2\cos (2\pi t) (1-{\hat {m}}_{x}) f_1 & ({\rm E1-E1~ \&~ E2-E2}) \\ 
2 i \sin (2\pi t) (1-{\hat {m}}_{x}) f_1 & ({\rm E1-E2}) 
\end{array} \right.
\label{scatter}
\end{eqnarray}

\noindent where we used the symmetry operations introduced above. The mirror ${\hat {m}}_{x}$ is such that (x,y,z)$\rightarrow$(-x,y,z). For the chosen reference frame its action on spherical tensors is given by\cite{ball}: ${\hat {m}}_{x} F^{(p)}_q = (-)^{P+p} F^{(p)}_{-q}$, where $P$ is the parity of the tensor (+1 for E1-E1 and E2-E2 tensors and $-$1 for E1-E2 tensors). The combined action of glide-plane and C$_3$-symmetry forbids any E1-E1 contributions, as expected, and  leaves us with only three possible terms: two come from the E1-E2 channel  ($F^{(2)}_0$ and $F^{(3)}_3 - F^{(3)}_{-3}$) and the other is of E2-E2 origin:  $F^{(4)}_3 - F^{(4)}_{-3}$, the one recognized in Ref. \onlinecite{carra}.

Their polarization and wavevector dependence shows that the signal is different from zero only in the $\sigma\pi$ channel and only for $T^{(2)}_0$ and $T^{(4)}_3 - T^{(4)}_{-3}$. In fact the $T^{(3)}_3 - T^{(3)}_{-3}$ term is proportional to $k_x-k'_x$ or $k_y-k'_y$, where $k$ and $k'$ are the incident and scattered wave vectors and $x$ and $y$ are orthogonal to the trigonal axis. In the geometry of the (00.3)$_h$, $k_x=k'_x$ and $k_y=k'_y$ at any azimuthal angle.

It is now easy to compare the theoretical azimuthal dependence around the momentum transfer $\hbar \vec{q}$ with the experimental one\cite{paolasini2}: the contribution of $T^{(2)}_0$ is constant with respect to the azimuthal angle $\phi$ around the $\bf {q}$-vector, while that of $T^{(4)}_3 - T^{(4)}_{-3}$ is three-fold periodical in the amplitude.\cite{carra} Note that time-reversal even quantities are real in the E2-E2 channel and imaginary in the E1-E2 channel.\cite{varma} Because of the imaginary unit in the E1-E2 term of Eq. (\ref{scatter}), both amplitudes are real and intefere. Thus, the global dependence in the scattered intensity is proportional to $(\alpha + \sin (3\phi))^2$, ie, a three-fold modulation of the six-fold periodic intensity. The constant $\alpha$ takes into account the relative weight of the radial matrix elements for the two irreducible tensors.

\begin{figure}
\vspace{-1cm}
\centerline{\epsfig{file=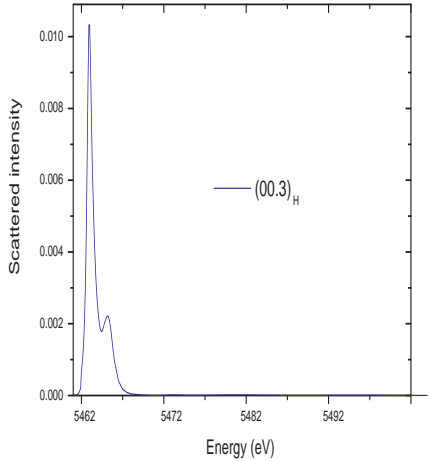,height=3.0in,width=2.5in}}
\vspace{-5pt}
\caption
{Numerical simulation for the energy scan of the (00.3)$_h$ reflection in corundum V$_2$O$_3$. Note the absence of any signal at the E1-E1 $4p$ conduction band energies. Here and in the following figures the intensity is proportional to the square of the effective number of scattering electrons.}
\label{enscan}
\end{figure}
\begin{figure}
\vspace{-1cm}
\centerline{\epsfig{file=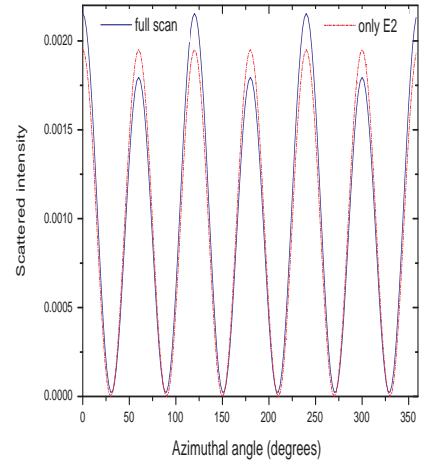,height=3.0in,width=2.5in}}
\vspace{-5pt}
\caption
{Numerical simulation for the azimuthal scan around the (00.3)$_h$ direction at vanadium pre K-edge (5465 eV). Also shown is the pure E2-E2 simulation in octahedral symmetry. In this last case the six-fold periodicity is clearly recovered.}
\label{azscan}
\end{figure}

In order to estimate such a constant  we have performed a numerical simulation based on the FDMNES package.\cite{yves2} The results are shown in Fig. \ref{enscan} for the energy scan and in Fig. \ref{azscan} for the azimuthal scan at the pre-edge energy (5465 eV). They both support our previous theoretical considerations: the energy scan shows no features at E1-E1 energies (as experimentally detected\cite{paolasini2}) while a structure exists around the 3d-energies (5465 eV). Also the azimuthal scan of this structure, shown in Fig. \ref{azscan} is compatible with the experimental data\cite{paolasini2} and makes clear its double-component origin. The possibility to evaluate separately the two contributions in the FDMNES program allows us to estimate the ratio between the constant E1-E2 signal and the maximum of E2-E2: $\alpha \simeq 0.05$. Note that this estimates includes the prefactors $\cos (2\pi t)$ and  $\sin (2\pi t)$, for E2-E2 and E1-E2. For $t=0.1537$ they are, respectively, $0.57$ and $0.82$.

A simple physical picture can help understanding the previous results. In the corundum system there are two main crystal fields that split the spherical degeneracy of the $3d$ electrons involved
in the resonant process: the octahedral crystal field is the biggest (10Dq $\simeq$ 2 eV) and is responsible of the splitting to doubly degenerate e$_g$ and triply degenerate t$_{2g}$ orbitals. The trigonal field, usually one order of magnitude less, further splits the t$_{2g}$ orbitals into doubly and singly degenerate levels. This field is made up of two contributions \cite{muto}: an axial field, of C$_{3v}$ symmetry, describing the shift of vanadium ions along the trigonal axis, and a ``twisted'' field of C$_3$ symmetry. This latter takes into account the small rotation of the oxygens around the trigonal axis, as well as the positions of the second-nearest-neighbors vanadium ions: both these latter features break the vertical mirror plane. In the absence of a trigonal splitting only $T^{(4)}_3 - T^{(4)}_{-3}$ contributes to the signal, as it is totally symmetric in an octahedral  potential. No inversion-breaking terms would be possible in this case. The detection of the term $T^{(2)}_0 \propto L_z\Omega_z$ is made possible only by the reduction of the local symmetry to C$_3$. Note that the allowed  $T^{(2)}_0$ tensor is  time-reversal even and inversion-odd. The identification with  $L_z\Omega_z$ follows from the fact that this latter is the only physical quantity detectable at the K-edge with these same properties.\cite{benoist,natoli} Notice also that if the point symmetry of vanadium ions were C$_{3v}$, as maintained by Carra {\it et al.},\cite{carra2} the term $L_z\Omega_z$ would be forbidden by symmetry, as it changes sign under the action of a vertical mirror plane and no modulation could be seen in the azimuthal scan.

All this has been tested by a numerical simulation where a ``straightened'' corundum cell has been used, with a perfect octahedral environment around each vanadium ion. In such a condition, there is no $T^{(2)}_0$ terms, as it should be. The E1-E2 signal, then, increases with the degree of ``twisting'' of the trigonal deformation of the oxygen octahedra.

\begin{figure}
\centerline{\epsfig{file=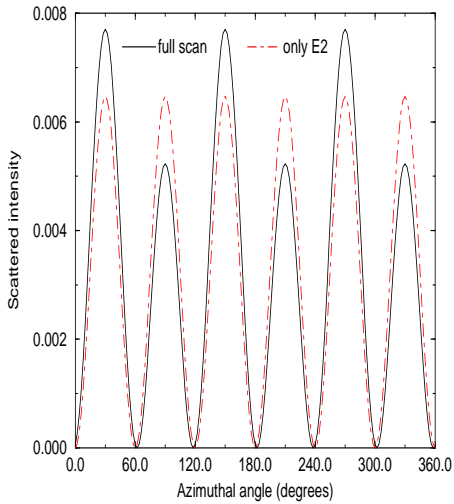,height=3.0in,width=2.5in}}
\vspace{-5pt}
\caption
{Same as Fig. 3 for $\alpha$-Fe$_2$O$_3$ (E=7105 eV).}
\label{azscan2}
\end{figure}

Similar results are obtained for the magnetic $\alpha$-Fe$_2$O$_3$, as shown in Fig. \ref{azscan2}. Neglecting for the moment magnetic contributions, we observe that in this case the ratio between two consecutive maxima is rather pronounced (about $3/2$), in keeping with the experimental findings.\cite{finkel} However it should be kept in mind that this ratio depends quite substantially on the photon energy, as numerically verified. If we now consider magnetic contributions, when the system is below the Morin temperature (T$_M\simeq$ 260 K) the magnetic moments are parallel to the trigonal axis and order in the sequency $\uparrow,\uparrow,\downarrow,\downarrow$ respectively for ions 1,2,3,4 in Fig. \ref{cell}. Thus the magnetic space group is again R${\overline {3}}$c and Eq. (\ref{scatter}) still holds. Yet, time-reversal is locally broken, and this allows in principle contributions from the tilded tensors, both in the E1-E2 and E2-E2 channels. From this point onward, tensor group analysis cannot be pushed further and we need quantitative calculations. Our numerical simulations give  a magnetic contribution about a factor $10^{-2}$ less than the non-magnetic one. This numerical result is in keeping with that of Ref. \onlinecite{finkel}, where the experimental data had been interpreted as non-magnetic, because the temperature dependence of the azimuthal periodicity was not affected by the transition through the Morin temperature. In fact, if the three-fold signal were magnetic, it should change drastically above T$_M$, where all the spins align perpendicularly to the trigonal axis. In this latter case, while the three-fold lattice symmetry is still preserved, the magnetic one is broken, and more tilded tensors components, with different periodicity, could in principle contribute.

Unfortunately, the small ratio of the magnetic to non-magnatic signal makes difficult, at present, a systematic study of the magnetic space group based on this kind of analysis for other corundum systems like Cr$_2$O$_3$ or Ti$_2$O$_3$.  Such a study would be interesting for many reasons: in the first case, in the light of the recent claim for magnetochirality,\cite{goulon2,dimatteo2} it could provide an independent verification. In the second case, because of an old controversy regarding the magnetoelectricity in Ti$_2$O$_3$,\cite{astrov} it could give more detailed information on the possible magnetic symmetries of the system. In all cases, the presence/absence of tilded tensors would be determined by their different polarization and wavevector dependence, as well as by the changes in Eq. (\ref{scatter}) due to the different magnetic space groups (in the case of Cr$_2$O$_3$, for example, both $\hat{I}$ and ${\hat{m}}_b$ appear with a minus sign for magnetic tensors). In  general for these and similar studies one should rely on better resolved spectra or on ``pure'' magnetic reflections. 

In conclusion, by a complete tensor analysis we have shown with a simple example that x-ray elastic resonant scattering and XNCD can be considered complementary tools for the detection of the chiral quantity $L_z\Omega_z$. In particular resonant diffraction, due to the possibility of adjusting phase factors, is the only technique that allows a detection of such a signal in systems where the inversion symmetry, though broken at a local level, is globally restored.

\end{multicols}


\begin{references}

\bibitem{paolasini}
L. Paolasini, C. Vettier, F. de Bergevin, F. Yakhou, D. Mannix, A. Stunault, W. Neubeck, M. Altarelli, M. Fabrizio, P.A. Metcalf and J.M. Honig,
 Phys. Rev. Lett. {\bf 82}, 4719 (1999)
\bibitem{goulon}
L. Alagna, T. Prosperi, S. Turchini, J. Goulon, A. Rogalev, C. Goulon-Ginet, C.R. Natoli, R.D. Peacock and B. Stewart, Phys. Rev. Lett. {\bf 80}, 4799 (1998);
J. Goulon, Ch. Goulon-Ginet, A. Rogalev, V. Gotte, C. Malgrange, Ch. Brouder and C.R. Natoli, J. Chem. Phys. {\bf 108}, 6394 (1998); C.R. Natoli, Ch. Brouder, Ph. Sainctavit, J. Goulon,  C. Goulon-Ginet, and A. Rogalev, Eur. Phys. J. B {\bf 4} 1-11 (1998)
\bibitem{sawatzky}
I.S. Elfimov, S.A. Skorikov, V.I. Anisimov, and G.A. Sawatzky, Phys. Rev. Lett. {\bf 88}, 015504 (2002)
\bibitem{benoist}
P. Carra, and R. Benoist, Phys. Rev. B, {\bf 62}, R7703 (2000)
\bibitem{finkel}
K.D. Finkelstein, Qun Shen, and S. Shastri, Phys. Rev. Lett. {\bf 69}, 1612 (1992)
\bibitem{paolasini2}
L. Paolasini, S. Di Matteo, C. Vettier, F. de Bergevin, A. Sollier, W. Neubeck, F. Yakhou, P.A. Metcalf, J.M. Honig, Jour. of Electr. Spectr. \& Related Phenomena, {\bf 120/1-3}, 1-10 (2001)
\bibitem{carra}
P. Carra, T. Thole, Rev. Mod. Phys. {\bf 66}, 1509, (1994)
\bibitem{ishida}
K. Ishida, et al. unpublished
\bibitem{dernier}
P.D. Dernier and M. Marezio, Phys. Rev. {\bf 2}, 3771 (1970)
\bibitem{natoli}
S. Di Matteo, C.R. Natoli, J. Synchrotron Rad. {\bf 9}, 9 (2002)
\bibitem{ball}
C.J. Ballhausen, {\it Introduction to Ligand Field Theory} 
(McGraw-Hill, New York, 1962)
\bibitem{varma}
S. Di Matteo, C.M. Varma, Phys. Rev. B {\bf 67}, 134502 (2003)
\bibitem{yves2}
Y. Joly, Phys. Rev. B {\bf 63}, 125120 (2001); The program can be freely downloaded at the web address: http://www-cristallo.grenoble.cnrs.fr/simulation
\bibitem{muto}
M. Muto, Y. Tanabe, T. Iizuka-Sakano, and E. Hanamura, Phys. Rev. B {\bf 57} 9586 (1998)
\bibitem{carra2}
P. Carra, A. Jerez, I. Marri, Phys. Rev. B {\bf 67} 045111 (2003)
\bibitem{goulon2}
J. Goulon, A. Rogalev, F. Wilhelm, C. Goulon-Ginet, P. Carra, D. Cabaret, and C. Brouder, Phys. Rev. Lett. {\bf 88}, 237401 (2002)
\bibitem{dimatteo2}
S. Di Matteo, C.R.Natoli, Phys. Rev. B {\bf 66}, 212413 (2002)
\bibitem{astrov}
B.I. Al'shin, and D.N. Astrov, Zh. Eksp. Theor. Fiz. {\bf 44}, 1195 (1963) [Sov. Phys. JETP {\bf 17}, 809 (1963)]; A.K. Agyei, J.L. Birman, J.Phys.:Cond. Mat., {\bf 2} (1990) 3007

\end{references}
\end{document}